\documentclass[
    ,final            
  ]
  {aipproc}

\layoutstyle{6x9}
\usepackage{amssymb}
\def\hi{\textsc{Hi} }

\begin{document}

\title{HI and Cosmology: What We Need To Know}

\classification{}
\keywords{Cosmology, Intergalactic Medium, Hydrogen, Reionization}

\author{Judd D. Bowman}{
  address={California Institute of Technology, Pasadena, CA 91125, USA}
  ,altaddress={Hubble Fellow}
}

\begin{abstract}
There are three distinct regimes in which radio observations of the redshifted 21 cm line of \hi can contribute directly to cosmology in unique ways.  The regimes are naturally divided by redshift, from high to low, into: inflationary physics, the Dark Ages and reionization, and galaxy evolution and Dark Energy.  Each measurement presents its own set of technical, theoretical, and observational challenges, making ``what we need to know'' not so much an astrophysical question at this early stage as a comprehensive experimental question. A wave of new pathfinder projects are exploring the fundamental aspects of what we need to know (and what we should expect to learn in the coming years) in order to achieve the goals of the Square Kilometer Array (SKA) and beyond.
\end{abstract}

\maketitle


\section{Introduction}

In the next decade, neutral hydrogen may become the ultimate cosmological probe.  The uniqueness of \hi stems from the expectation that, in principle, its presence can be detected, and its properties characterized, during every epoch of cosmological history.  Immediately following recombination, during a period commonly called the Dark Ages, neutral hydrogen was the dominant form of baryonic matter, existing everywhere as a diffuse, pervasive gas.  Eventually, though, gravitational instabilities in this gas, induced by primordial matter density fluctuations, led to the formation of the first stars, galaxies, and quasars.  New radiative processes from these collapsed objects exerted, for the first time, substantial feedback on the diffuse \hi that remained in the intergalactic medium (IGM), and vast regions of neutral hydrogen became ionized.  After some time, only localized pockets of \hi were left in, and around, galaxies.  These galactic \hi regions persist to the present day and play active roles in the evolution of galaxies.  The transition period between these two primary eras---one characterized by large amounts of diffuse \hi in the IGM, and the other by small regions of localized \hi in galaxies---is known as the epoch of reionization (EOR).  It is constrained to occur between $6\lesssim z \lesssim15$ by existing observations, although with considerable uncertainty in the specific details of the transition.

Unlike the Lyman series of electronic transition lines, the \hi 21~cm hyperfine spin-flip transition line does not suffer an optical depth problem at high redshift and the IGM remains optically thin over all redshifts.  The redshifted 21~cm line, whether in emission or absorption, is weak (particularly compared to other sources in the radio sky), however, and detecting and characterizing 21~cm emission or absorption at cosmological distances is very difficult.  All-sky surveys to detect galaxies in 21~cm emission, such as the Arecibo Legacy Fast ALFA Survey (ALFALFA) \citep{2005AJ....130.2598G}, have reached to only $z\lesssim0.1$.  And the highest confirmed detections of 21~cm absorption in damped Lyman-$\alpha$ systems are at only $z\lesssim3$ \citep{2007MNRAS.375.1528K}.  Overcoming the sensitivity limits needed to thoroughly characterize 21~cm emission and absorption from $0 < z \lesssim 15$ is one of the primary motivations for the Square Kilometer Array\footnote{http://www.skatelescope.org} (SKA).  But moving from the current paradigm, where detections of discrete objects are difficult even at the lowest redshifts, to a large, successful \hi science program with the SKA is a path filled with significant uncertainty and numerous challenges.

\section{\hi Cosmology}

The redshifted 21~cm radiation targeted by the SKA and other upcoming experiments falls in the frequency range $10 \lesssim \nu < 1420$~MHz for \hi below $z\lesssim200$.  The potential contributions of \hi to cosmology can be roughly divided into three regimes defined by frequency, or redshift, throughout this range.  The three regimes are (see the recent review at \cite{2006PhR...433..181F} for details):

\textit{Inflationary Physics [$30 < z < 200, \; \; \; 46 > \nu > 7$~MHz]}.  At very high redshifts, \hi is expected to be a clean tracer of baryonic matter.  As such, constraining the anisotropic fluctuations of the redshifted 21 cm signal could probe the matter power spectrum at very small scales of order $\ell \gtrsim 10^4$ to $10^6$ that are unattainable with CMB anisotropy measurements or galaxy surveys.  These observations would be able to constrain perturbations to the primordial power spectrum and spatial curvature (including parameters such as $n_s$ and $\alpha_s$), neutrino masses, non-Gaussianity, and inflationary models.  At such high redshifts, the perturbations in the matter density fluctuations should be linear and relatively easy to interpret. In addition, the anisotropy power spectrum from \hi is three-dimensional since the signal is a spectral line (as opposed to the two-dimensional CMB arising from continuum emission), and thus contains more Fourier modes than the CMB, potentially providing less cosmic sample variance and better cosmological parameter constraints.  The process of baryon collapse into dark matter gravitational potential wells following recombination also should be seen to unfold with redshift in \hi observations during this era.

\textit{Dark Ages and Reionization [$6 < z < 30, \; \; \; 203 > \nu > 46$~MHz]}.  Redshifted 21 cm emission and absorption offers truly unparalleled views into the evolution of the IGM during the crucial times associated with the formation of the first stars, galaxies, and quasars.  Measurements of both the mean (global) redshifted 21 cm brightness temperature and the fluctuation power spectrum (both with characteristic amplitudes of order 10~mK) should yield the spin and kinetic temperature histories of the IGM and the reionization history.  Indirectly, these measurements probe the early star formation history and the nature of the luminous sources responsible for ionizing photons.  Cross-correlation of (even low signal-to-noise) \hi maps with CMB maps or planned high-redshift galaxy surveys could add additional insight into the processes responsible for reionization.  High signal-to-noise maps during this era could yield images of Stromgren spheres (ionized bubbles) around individual quasars and tomographic maps of the IGM.  Together, these measurements could determine the abundance of mini-halos during the Dark Ages and weakly constrain magnetic fields in the IGM.  Recent efforts suggest that cosmological parameter estimation solved simultaneously with parameterized reionization models in measured power spectra over a range of redshifts could yield valuable improvements to the parameter limits achievable even with Planck (in conjunction with other cosmological data sets).  This process exploits the three-dimensional nature of the \hi power spectrum to separate cosmological and astrophysical contributions using velocity-field effects inherent in redshift-space distance measurements.  Although not a primary driver of \hi cosmology, characterization of 21~cm absorption by \hi along the line-of-site toward bright, high-redshift sources such as active galactic nuclei (AGN), star-forming galaxies, and gamma-ray bursts (GRBs) should be possible during this era, as well, with the SKA.  And finally, \hi maps and power spectra from this era could provide source planes for weak lensing studies of the matter distribution at lower redshifts.

\textit{Galaxy Evolution and Dark Energy [$0 < z < 6, \; \; \; 1420 > \nu > 203$~MHz]}.  Following reionization, localized galactic clumps of \hi can be detected individually and cataloged (with $\gtrsim 10^8$ entries expected for the SKA), or characterized using the same diffuse mapping approaches needed to exploit \hi in the IGM during and before reionization.  Both of these approaches could lead to very accurate estimates of the matter power spectrum suitable for characterizing baryon acoustic oscillations (BAOs) and constraining the nature of Dark Energy.  In addition, $\Omega_{HI}(z)$ should be well constrained.  These measurements are also very valuable for studying galaxy evolution, including, in the case of diffuse maps, through cross-correlations with galaxy surveys divided by galaxy type or environment and redshift.  And, as already demonstrated below $z\lesssim3$, 21~cm absorption toward by \hi in damped Lyman-$\alpha$ systems and other objects along the path to bright, discrete sources is an existing technique that will only expand with the SKA and other new instruments.

\section{Challenges and Unknowns}

The potential rewards of \hi cosmology are compelling, but the challenges are substantial.  Experiments operating in the frequency range needed for redshifted 21 cm measurements face three general categories of hurdles:

\textit{Radio Frequency Interference}.  The frequencies that will be targeted are commonly used for television, FM radio, satellite, cellular phone, and other communications transmissions. The candidate sites for the SKA, in Western Australia and South Africa, are in remote locations to avoid these radio communications, but it will almost certainly be impossible to completely eliminate all of the sources of interference.  No telescope has ever reached the sensitivity levels planned for \hi observations with the SKA and it is unknown just how deep the integrations will be able to go before reaching a limiting interference floor.

\textit{Ionosphere}.  Turbulence in the Earth's ionosphere refracts low frequency radio waves (particularly relevant below $\sim300$~MHz) and causes distortions in the apparent location and magnitude of signals originating from above the ionosphere.  Correcting for these distortions over wide fields-of-view and very long baselines should be possible \cite{2004SPIE.5489..180C}, but has never been demonstrated under the full set of conditions that will exist for the most ambitious planned high-redshift 21~cm observations.

\textit{Astrophysical emission}.  Astrophysical foregrounds are $10^3$ to $10^9$ times brighter than the predicted $\sim1$ to 100~mK redshifted 21~cm signal everywhere on the sky.  Galactic synchrotron radiation tends to be the dominate contribution over most of the frequency range, but extragalactic continuum point sources are also especially strong and numerous. Galactic radio recombination lines (RRLs) and free-free emission from electrons in both the Galaxy and the IGM should be present at roughly the level of the expected signal. Synchrotron radiation from high Galactic latitudes has frequency-dependent intensity of approximately $T_{sky}(\nu) \approx 250 (\nu/150$~MHz$)^{-2.6}$~K.  It is also linearly polarized at the percent level ($\sim1$~K) and exhibits a high degree of structure due to Faraday rotation by the interstellar medium.  The astrophysical foregrounds set not only stringent instrumental calibration and data analysis performance requirements, but dictate the system temperatures of the instruments since $T_{sky}$ is well above the typical receiver temperatures ($T_{rcv}\approx50$~K) of radio instruments, particularly for \hi measurements targeting redshifts above $z\gtrsim6$ ($\nu\lesssim200$~MHz).  Therefore, the foreground sky temperature governs the collecting area needed for the \hi science drivers of the SKA and other experiments and, in general, forces the arrays to be very large.  For this reason, it is unlikely that the inflationary physics regime of \hi cosmology above $z\gtrsim30$ will be accessible for many years since, in this regime, $T_{sys} \approx T_{sky} \gtrsim 5000$~K.

Mitigating the astrophysical foregrounds is anticipated to require a multi-stage foreground subtraction effort that exploits the power-law-like spectral smoothness of all the foreground categories (except RRLs) by fitting and subtracting low-order polynomials along each line-of-sight in an interferometric data cube.  First, deconvolution of bright continuum sources within the target field will be performed using techniques similar to iterative ``pealing''.  This will be followed by subtraction of the predicted contributions (due to calibration uncertainties and other instrumental paths) of both discrete and diffuse foregrounds from outside the primary target fields using global sky models.  Interferometric measurements will still be dominated after this process by faint confusion-level sources that are mixed by the array beam sidelobes.  Recent efforts suggest that, by careful instrumental design and data processing, the contribution of these sources can be subtracted even in ``dirty'' sky maps.  In principle, this multi-stage process should be sufficient to reveal the redshifted 21 cm signal.  However, it is also anticipated that polarized foregrounds will leak into the desired intensity measurement through mis-calibrations.  Techniques for mitigating this contamination are an area of ongoing development, but approaches based on separation of signal and foreground through rotation measure synthesis appear encouraging.  In addition, model errors and fitting uncertainties in the foreground subtraction process could manifest themselves at this stage (if not earlier) and methods for identifying and ameliorating these effects have proposed \cite{2006ApJ...648..767M}.  The last step will be data consistency cross-checks to confirm that foregrounds have been separated to the required precision.  These could include cross-correlation of different Stokes maps, inspection of noise correlation matrices, comparison of measurements at different redshifts (such as at $z=8$ when a reionization signature should be present and at $z=5$ when virtually no redshifted 21 cm contribution should exist), or cross-correlation of \hi maps and galaxy surveys.

All of the terrestrial and astrophysical foregrounds will severely complicate observations of weak redshifted 21~cm emission in the high-redshift IGM, as well as in galaxies at lower redshifts.  With little current empirical knowledge about their detailed properties to guide the development of redshifted 21~cm experiments and with a large range of instrumental approaches still under consideration for the SKA, learning more about the foregrounds and the consequences of instrumental design choices on mitigating them is at the core of ``what we need to know'' to unleash the full cosmological potential of neutral hydrogen.  Thus, to a large degree, what we need to know to enable \hi cosmology is fundamentally intertwined with questions regarding what we need to know to successfully develop the SKA and other new redshifted 21~cm radio arrays.

\section{Experimental Parameter Space}

Addressing these questions and hurdles now rests firmly on a wave of new experiments, including the Murchison Widefield Array\footnote{http://www.haystack.mit.edu/ast/arrays/mwa} (MWA) \cite{2005SPIE.5901..124S, 2005ApJ...634..715W, 2006ApJ...638...20B, 2006ApJ...653..815M, 2007ApJ...661....1B, 2007AJ....133.1505B}, the Precision Array to Probe the Epoch of Reionization (PAPER), and the Low Frequency Array\footnote{http://www.lofar.org} (LOFAR) \cite{2005MNRAS.360L..64Z, 2006MNRAS.369L..66V, 2006MNRAS.373..623R, 2006ApJ...653..815M}, as well as pilot projects on existing facilities such as the Giant Metre-wave Radio Telescope (GMRT) and with existing low-redshift data sets from the \hi Parkes All Sky Survey (HIPASS) \cite{2001MNRAS.322..486B, 2007MNRAS.378..301B, 2008arXiv0802.3239P} and other observations.  In addition to these efforts, the Alan Telescope Array\footnote{http://ral.berkeley.edu/ata} (ATA) and the Expanded Very Large Array\footnote{http://www.aoc.nrao.edu/evla} (EVLA) will, in many ways, act as SKA technology pathfinders as they become increasingly operational over the next few years, as will several explicit SKA precursor projects including the Australian SKA Precursor\footnote{http://www.atnf.csiro.au/projects/askap} (ASKAP) and the South African MeerKAT\footnote{http://www.ska.ac.za/meerkat}.  In addition, new projects not on the pathway to the SKA are tackling measurements of the mean redshifted 21~cm signal as a function of redshift.  These include the Compact Reionization Experiment (CoRE) and the Experiment to Detect the Global EOR Signature\footnote{http://www.haystack.mit.edu/ast/arrays/Edges} (EDGES) \cite{2008ApJ...676....1B}.

The landscape of these precursor-level redshifted 21~cm projects will undoubtedly evolve on the way to the SKA, but the process as a whole will greatly benefit the final design, development, and operation of the SKA and other future experiments as lessons from the successes and failures of these early trailblazers become incorporated into subsequent efforts.  Indeed, the current environment mirrors the history of the experimental cosmic microwave background (CMB) community throughout much of the 1980s and 1990s \cite{2003PhRvD..68l3001W} along the way to the successful Wilkinson Microwave Anisotropy Probe (WMAP).  It is interesting to note from the CMB experience that one technology does not necessarily emerge as the only alternative and that, for different specific goals within the same broad science objectives, multiple technologies can be exploited fully.  Early redshifted 21~cm projects that are no longer active, including the VLA EOR Extension Program\footnote{http://www.cfa.harvard.edu/dawn} (VEEP) and the Primeval Structure Telescope\footnote{http://web.phys.cmu.edu/~past} (PAST) \cite{2004astro.ph..4083P}---also called the 21~cm Array (21CMA)---and preliminary experiments with the GMRT \cite{2008MNRAS.tmp..301A} have already helped to refine the course to \hi cosmology by demonstrating that significant foreground and technical challenges do in fact exist and are not easily overcome by existing facilities.

\section{Putting It All Together}

The primary goal for the growing \hi cosmology community over the next five years is to find out if carefully designed experiments and well devised analysis techniques can mitigate the effects of the terrestrial and astrophysical foregrounds sufficiently well to enable the detection of redshifted 21 cm emission over large volumes of the universe.  In order to accomplish this task, new levels of instrumental calibration and ionospheric calibration will need to be achieved and robustly characterized, the statistics of faint source populations will need to be documented since even confusion noise will be well above the \hi signal in most experiments, the spectral coherence function of foregrounds will need to be confirmed on small scales in order to permit the foreground subtraction techniques described above, and techniques for dealing with polarized foregrounds will need to be implemented in new ways. When all of these issues have been successfully addressed and robust detections of \hi during and after the reionization epoch have been achieved, the field will transition from its preliminary \textit{detection} phase to a \textit{characterization} period leading to the SKA and beyond.  During this time, attention will shift to ensuring that basic theoretical expectations match with observations and then building a solid model-based parametrization for connecting theory and observation through a small number of core processes.

As a set of goals to focus efforts, it is reasonable to assert that by the end of the next decade (2020) we should have robustly explored the low-frequency radio and digital signal processing technological parameter spaces, have multiple independent detections of anisotropy spectra at $z > 6$, as well as at $z<6$, have significantly constrained the mean (global) spin, kinetic, and ionization histories of the IGM, and know if HI cosmology ultimately needs to go to the moon (to overcome RFI or the ionosphere).  If all of this has been accomplished, we will truly need an SKA-class instrument to make more progress.



%


\begin{theacknowledgments}
  JDB is supported by NASA through Hubble Fellowship grant HF-01205.01-A awarded by the Space Telescope Science Institute, which is operated by the Association of Universities for Research in Astronomy, Inc., for NASA, under contract NAS 5-26555.
\end{theacknowledgments}



\end{document}